\documentclass[10pt,twocolumn]{article}

\usepackage{graphicx}
\usepackage{indentfirst,csquotes} 
\usepackage{amssymb,amsthm,amsmath}
\usepackage{xcolor,paralist,hyperref,titlesec,fancyhdr}
\usepackage{siunitx}
\usepackage[numbers,sort&compress]{natbib}
\usepackage{booktabs}
\usepackage{float}
\usepackage{braket}
\usepackage{multirow}
\usepackage{dblfloatfix}
\usepackage{etoolbox}
\usepackage{balance}
\makeatletter
\AtBeginEnvironment{thebibliography}{%
  \clubpenalty=150 \@clubpenalty=150 \widowpenalty=150 \interlinepenalty=0 }
\makeatother

\newcommand{\br}{\bottomrule}
\newcommand{\mr}{\midrule}

\usepackage[
  letterpaper,
  left=0.75in,right=0.75in,
  top=0.80in,bottom=0.80in,
  headheight=13.6pt,
  headsep=2pt,
  footskip=14pt
]{geometry}

\titleformat{\section}{\normalfont\large\bfseries}{\thesection}{1em}{}
\titlespacing*{\section}{0pt}{1ex}{1ex}

\hypersetup{colorlinks=true, linkcolor=black, filecolor=black, urlcolor=black, citecolor=blue}

\titleformat{\section}
  {\centering\normalfont\normalsize\bfseries}
  {\thesection.}
  {0.6em}
  {\MakeUppercase}
\titlespacing*{\section}{0pt}{3.0ex}{0.8ex}

\titleformat{\subsection}
  {\centering\normalfont\normalsize\bfseries}
  {\thesubsection.}
  {0.6em}
  {}
\titlespacing*{\subsection}{0pt}{2.5ex}{0.8ex}

\makeatletter
\newcommand{\captionsize}{9.0}
\newcommand{\captionbaselineskip}{10.0}
\renewcommand\fnum@figure{Figure~\thefigure.}
\renewcommand\fnum@table {Table~\thetable.}
\long\def\@makecaption#1#2{%
  \vskip\abovecaptionskip
  \begingroup
  \rmfamily\fontsize{\captionsize}{\captionbaselineskip}\selectfont
  \sbox\@tempboxa{#1~#2}%
  \ifdim\wd\@tempboxa>\hsize
    \noindent #1~#2\par
  \else
    \hb@xt@\hsize{\hfil\box\@tempboxa\hfil}%
  \fi
  \endgroup
  \vskip\belowcaptionskip
}
\makeatother

\makeatletter
\let\oldtable\table
\let\endoldtable\endtable
\renewenvironment{table}[1][tbp]{
  \oldtable[#1]
  \footnotesize
}{\endoldtable}
\makeatother

\begin{document}

\twocolumn[
\begin{@twocolumnfalse}
\begin{center}
{\large\bfseries The Natural Formulation of Cubic-Phase Dynamics in the Linear Potential\par}
\vspace{0.8ex}
{\normalsize Maximilian L.~D.~D.~Pellner$^{1,2}$ and Georgi Gary Rozenman$^{2}$\par}

\vspace{0.4ex}
{\small
$^{1}$Fakultät für Physik, Ludwig-Maximilians-Universität, 80799 München, Germany\\
$^{2}$Department of Mathematics, Massachusetts Institute of Technology, Cambridge, Massachusetts 02139, USA\\

\today
\par}
\end{center}

\vspace{-2.8ex}
\begin{center}
\begin{minipage}{0.80\textwidth}
\small\hspace*{0.3cm}
A quantum wave packet in a linear potential, i.e.\ a constant force such as gravity, accumulates a cubic-in-time phase, universal across Schrödinger-type platforms.
The natural formulation of the cubic phase dynamics follows from their canonical realization, the Airy eigenstates.
The eigenforce, the potential gradient for which the Airy profile is stationary and hence the intrinsic force behind its known self-acceleration absent any applied force, derives from the eigenvalue problem and characterizes the state uniquely up to a rigid translation.
Alongside the applied force, it underlies the analytically closed form of these dynamics, comprising three known contributions, intrinsic, force-induced, and a cross term.
The force-induced one alone is shape-independent, whereas the Airy eigenstate renders the other two non-dispersing.
The cubic coefficient factorizes into two Newtonian force balances, vanishing at the stationary eigenstate and at a non-trivial zero where the phase cancels while the packet keeps accelerating.
This exposes the eigenforce as an effective antagonist to the applied force, not only in the caustic's self-acceleration but here also within the phase, while leaving the centroid untouched by Ehrenfest's theorem.
Spatially uniform within each packet, it escapes direct measurement, accessible only as the relative phase of two colliding packets, each in its own potential.
The relative cubic coefficient is symmetry-forbidden for identical packets and activated by preparation asymmetry.
Simulated interference of two truncated Airy packets demonstrates the extraction.
Spanning ultracold atoms, paraxial optics, and surface-gravity water waves, it provides a platform-independent probe of the cubic phase and, through its force dependence, of the constant force itself.

\end{minipage}
\end{center}
\vspace{1.2ex}

\end{@twocolumnfalse}
]

\thispagestyle{empty}

\section{Introduction}

\begin{figure*}[t]
  \centering
\includegraphics[width=\textwidth,trim=220 1750 170 200,clip]{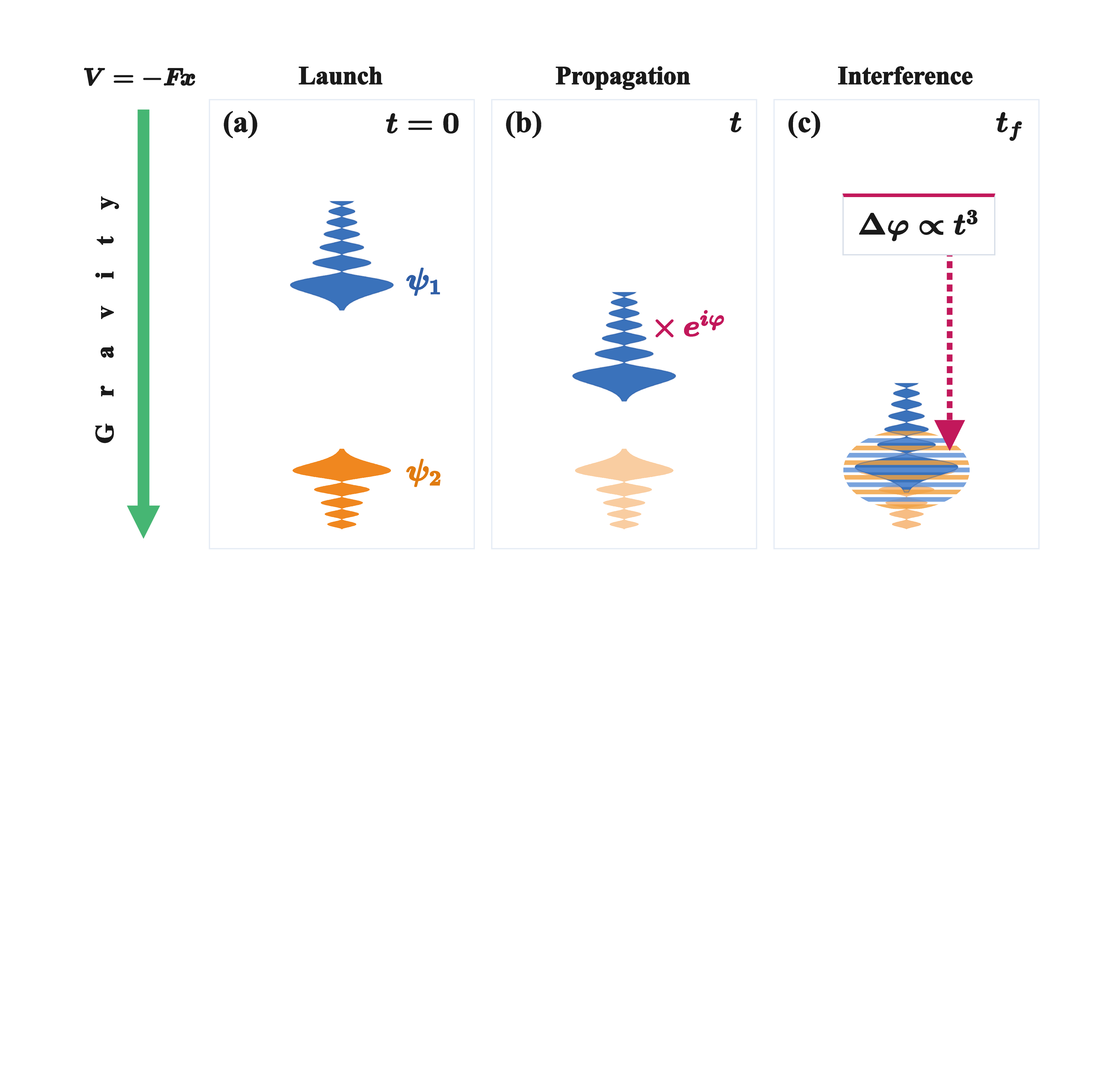}
  \caption{\textbf{Propagation and interference of Airy wave packets in a linear potential.} Conceptual evolution of a
  falling Airy wave packet in a linear potential $V=-Fx$ (e.g.\ gravity).
  \emph{(a)} The Airy wave function $\psi_1$ is launched above a
  stationary Airy function $\psi_2$. \emph{(b)} Under the constant force $F$ it
  accelerates downward, accruing a cubic-in-time phase. \emph{(c)} The two packets
  overlap and interfere, imprinting the relative phase on the fringe density,
  $\rho\propto\cos\Delta\varphi$, whose cubic-in-time part is
  $\Delta\varphi\propto t^{3}$. The width of each packet encodes its density
  $|\psi(x)|^{2}$ along the propagation axis $x$ (mirrored for symmetry), not a
  transverse coordinate.}
  \label{fig:concept}
\end{figure*}

Quantum mechanics, contrary to its classical counterpart, ascribes a wave nature to matter,
which motivates the definition of a complex wavefunction encoding these undulatory dynamics and carrying a phase with no analogue in a classical trajectory.
This phase governs how quantum amplitudes combine, yet a phase shared by an entire isolated wavefunction leaves the probability density untouched~\cite{born1926}, thereby escaping direct measurement.
What remains physical is the relative phase between superposed components, which reshapes the probability as soon as those components are brought to interference.
Interference thus provides insight into the quantum phase, and recent progress in atomic, molecular and optical physics, 
from the realization of Bose-Einstein condensates~\cite{ketterle1995,cornell1995} to the routine control of coherent matter waves~\cite{Cronin2009},
has further advanced this established field of phase measurement.

In parallel, the linear potential – the abstract formulation of a constant force, a ubiquitous concept, e.g. a uniform electric field, locally uniform gravity, or a uniformly accelerated frame
– is fundamental in two respects: it ranks among the few exactly solvable quantum systems, and represents the first order reduction of any smooth potential, making it the generic local approximation of arbitrary potentials.
In particular, this very approximation underlies the semiclassical Airy description near a classical turning point within the WKB approximation~\cite{landau_nonrelativistic,berrymount1972}.
A quantum particle evolving in such a potential keeps its centroid on the classical trajectory – in accordance with Ehrenfest's theorem, so that much of the distinctly wave-mechanical content is carried by the phase it accumulates.

The eigenstates of the linear potential are the Airy functions – a family of non-normalizable solutions in their ideal form, 
characterized by remarkable properties, i.e. self-acceleration of the profile despite the absence of any force and dispersionless propagation~\cite{berrybalasz1979}, as well as self-healing of the wave shape after partial obstruction~\cite{broky2008}.
Their physically admissible, truncated, finite-energy realizations retain these behaviours over finite times, generated as truncated Airy beams in optics~\cite{siviloglou2007ol,siviloglou2007prl,efremidis2019overview} and observed in ultracold atoms as a standing $^{87}$Rb matter wave off a linear magnetic potential~\cite{koehl2001}.

Part of the phase dynamics responsible for these features is a spatially uniform, cubic-in-time phase that a freely propagating Airy profile acquires~\cite{berrybalasz1979},
an intrinsic property rooted in its role as the eigenstate of the linear potential.
Adding a linear potential introduces a second cubic-in-time, now force-induced contribution~\cite{kennard1927,kennard1929,zimmermann2017}. Their interplay adds a third, cross contribution, made explicit by the derivation below.
These cubic dynamics have since been resolved experimentally, the force-induced contribution isolated in a Stern–Gerlach interferometer~\cite{amit2019} and, in surface-gravity water waves, reconstructed in both amplitude and cubic-in-time phase for Airy and Gaussian packets and compared with the single-packet cubic coefficient in closed form~\cite{rozenman2019amplitude,rozenman2020observation,rozenman2021projectile}.
Most recently, the phase dynamics of interfering Airy Bose–Einstein condensates in free space have been studied numerically~\cite{pellner2026phase}.

\begin{figure*}[t]
  \centering
  \includegraphics[width=\textwidth,trim=5 0 0 0,clip]{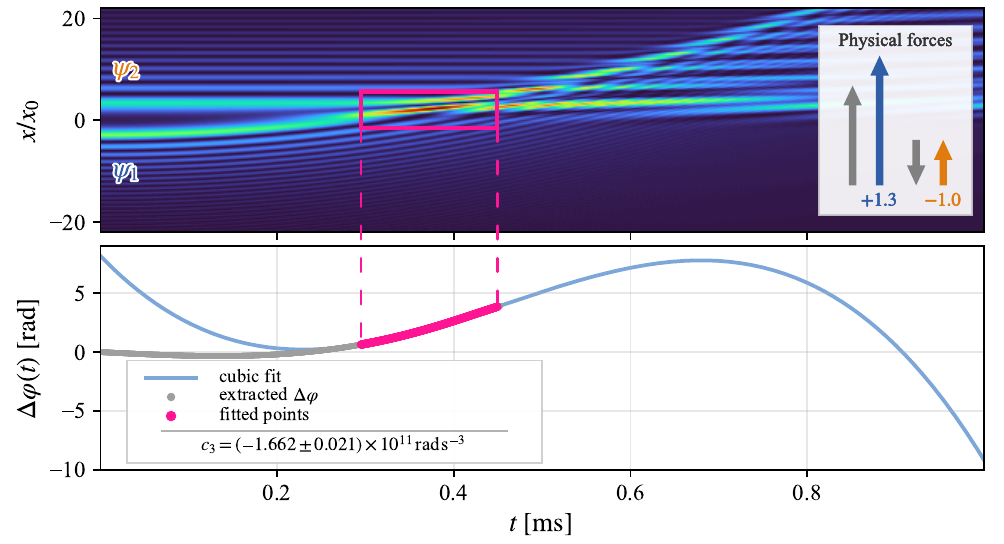}
  \caption{\textbf{Extraction of the relative cubic phase.} Simulated
  Airy--Airy collision ($^{87}$Rb) in the linear potential $V=+Fx$, propagated by a
  split-step Fourier integrator. \emph{Top:} probability density
  $|\psi(x,t)|^{2}$ against position in units of the Airy length $x_0$, the density itself shown qualitatively in colour. The pink rectangle marks
  the main interference fringe, and the dashed connectors link its lower corners
  to the outer points of the fit window. \emph{Bottom:} relative phase
  $\Delta\varphi(t)$ recovered by heterodyne demodulation (grey: extracted
  $\Delta\varphi$, pink: the fitted points) with the cubic least-squares fit (blue). The
  phase is extracted only over the first interference fringe, and points
  beyond it, where the position-dependent phase is no longer reliably concatenated,
  are omitted~\cite{note_phase_cutoff}.
  Packet~2
  is prepared as an approximately stationary Airy eigenstate, for which its single-packet coefficient vanishes approximately, so the relative
  cubic coefficient~(Eq.~\ref{eq:c3rel}) reduces to the single-packet value of the
  accelerating packet~1. In contrast to the gravitational, downward launch
  sketched in Fig.~\ref{fig:concept}, packet~1 here accelerates upward onto the
  stationary packet~2, the direction set by the sign of its applied force. The
  roles and the relative cubic phase carry over unchanged. \emph{Right inset:}
  per-packet forces. For each packet (coloured as in the density panel, packet~1 blue,
  packet~2 orange) the coloured arrow is the applied force and the grey arrow its
  eigenforce, drawn on a common scale and signed by the arrowhead end, the number
  giving the ratio $F_{\mathrm{phys},i}/F_{\mathrm{eigen},i}$ ($-1$ at an
  eigenstate). The two
  mirror packets carry opposite eigenforces, and packet~2, whose applied force
  opposes its eigenforce, sits on the eigenstate line.
  The fit returns
  $c_3=(-1.662\pm0.021)\times10^{11}\,\mathrm{rad\,s^{-3}}$ against the predicted
  $-1.662\times10^{11}\,\mathrm{rad\,s^{-3}}$, a deviation of $0.01\%$ for the calibrated
  estimate at this parameter configuration, realizing one
  point of the cubic-phase landscape~(Fig.~\ref{fig:landscape}). The quoted
  uncertainty is the ordinary least-squares standard error of $c_3$. Parameters:
  width scalings $\beta_1=1.0$, $\beta_2=1.3$, which set the Airy length per
  packet as $x_{0,i}=\beta_i x_0$, applied forces
  $F_{\mathrm{phys},1}=1.3\,F_{\mathrm{eigen},1}$,
  $F_{\mathrm{phys},2}=-F_{\mathrm{eigen},2}$,
  truncation $a=0.05$, fit window $[0.295,0.449]\,\mathrm{ms}$.}
  \label{fig:hero}
\end{figure*}

The natural formulation of these cubic phase dynamics rests on the eigenforce of their canonical realization, the Airy eigenstates. Fixed by the stationarity condition (Sec.~\ref{sec:model}), it characterizes them uniquely up to a rigid translation and drives the Airy self-acceleration in the absence of any applied force.
The cubic phase consists of three contributions, the force-induced one originating from Kennard~\cite{kennard1927,kennard1929}, the Airy-intrinsic one, and a cross term, all three already implicit in the general-force solution of Berry--Balazs~\cite{berrybalasz1979}. Of these, the first is shared by arbitrary wave shapes (Sec.~\ref{sec:universality}).
In the eigenforce and the applied force the cubic coefficient factorizes, exposing the force structure it encodes and vanishing twice: once where the packet is at rest, and once where it keeps accelerating. In parallel to the known caustic dynamics~\cite{berrybalasz1979}, the two forces hence enter the phase as a Newtonian force balance, although the eigenforce is effective rather than real and belongs to the state and not to the Hamiltonian.
Carried into the relative phase of two colliding packets, in general each in its own potential, this structure becomes a directly measurable observable (Fig.~\ref{fig:hero}), demonstrated here on simulated collisions of truncated Airy packets (Sec.~\ref{sec:demonstration}).

Beyond the genuinely quantum case, the same phase dynamics govern analogue platforms whose wave fields obey a Schrödinger-type equation of identical structure, among them paraxial optics and surface-gravity water waves~\cite{rozenman2019quantum} (Table~\ref{tab:platforms}).
These classical systems provide analogue platforms, making the quantum phase dynamics directly accessible to experiment – even where the underlying field is itself classical.
Resting solely on the structure of the Schrödinger equation, the general cubic phase dynamics characterized here are universal across all such platforms — each a distinct realization of one and the same description — and extend to further Schrödinger-type settings beyond those listed, from electron waves~\cite{volochbloch2013} and neutron beams~\cite{sarenac2024neutron} to acoustic waves~\cite{zhang2014acoustic}, surface plasmons~\cite{minovich2011plasmon}, and linear photonic lattices~\cite{christodoulides2003lattices,lucic2013airy}.

\begin{table*}[t]
  \centering
  \small
  \setlength{\tabcolsep}{14pt}
  \renewcommand{\arraystretch}{1.7}
  \begin{tabular}{@{}clcc@{}}
    \toprule
     & Platform & Free Schrödinger-type equation & Linear potential \\
    \mr
    \raisebox{-0.45\height}{\includegraphics[height=20mm,trim=0 100 0 100,clip]{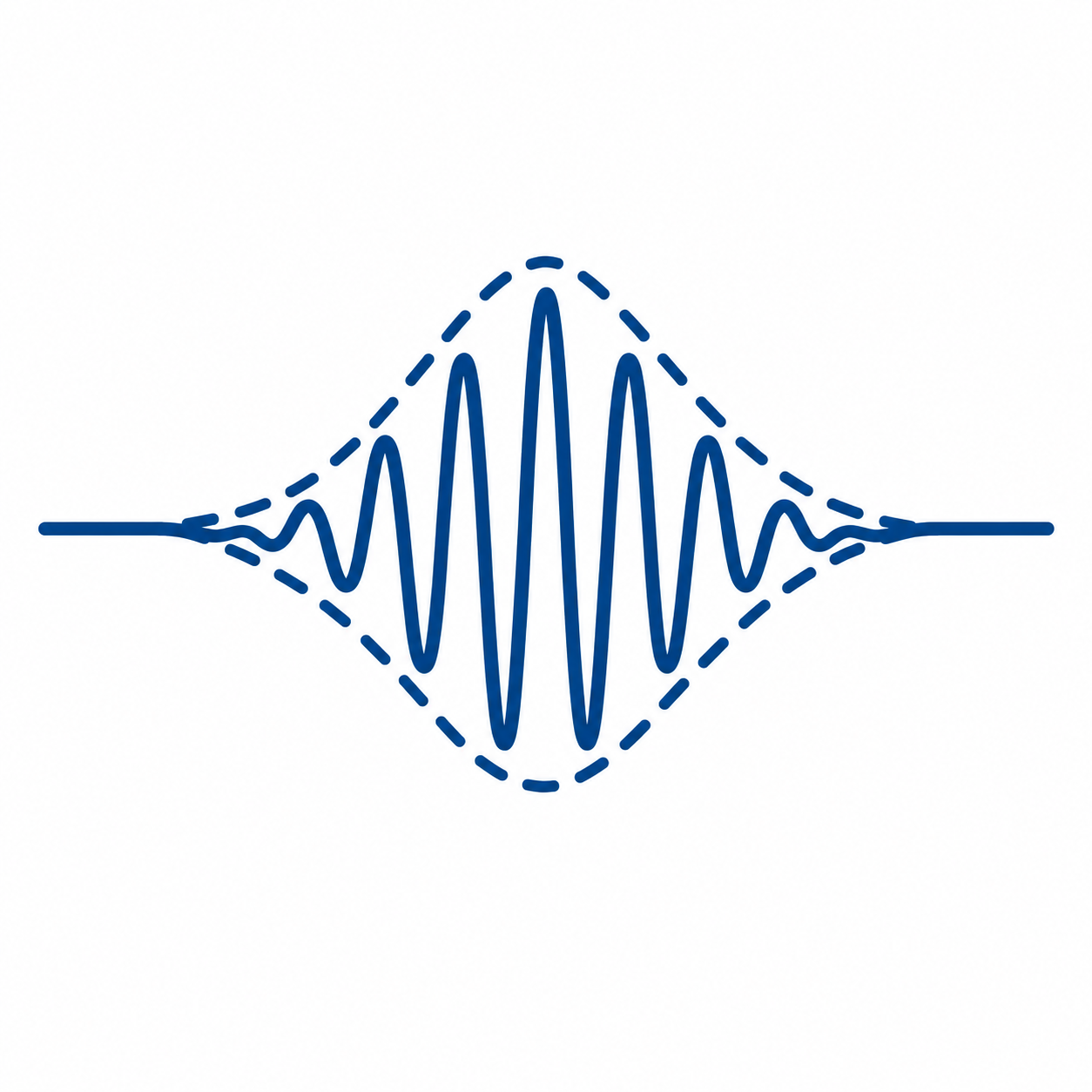}}
      & Quantum mechanics      & $\displaystyle i\hbar\,\partial_t\Psi + \tfrac{\hbar^2}{2m}\,\partial_x^2\Psi = 0$ & $V=-Fx$ \\
    \raisebox{-0.45\height}{\includegraphics[height=20mm,trim=0 100 0 100,clip]{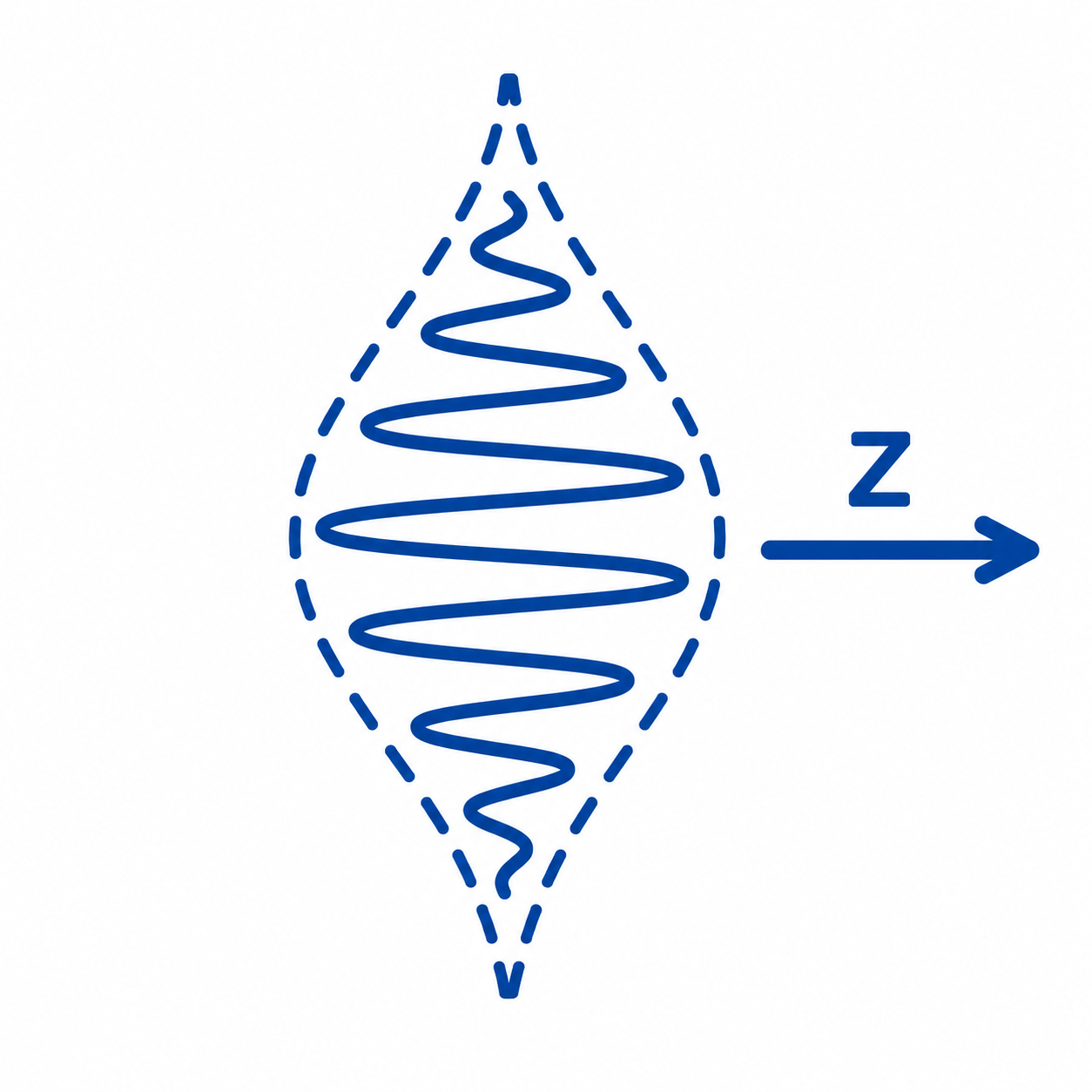}}
      & Paraxial optics        & $\displaystyle i\,\partial_z U + \tfrac{1}{2k}\,\partial_x^2 U = 0$               & $V=-Fx$ \\
    \raisebox{-0.45\height}{\includegraphics[height=20mm,trim=340 455 107 180,clip]{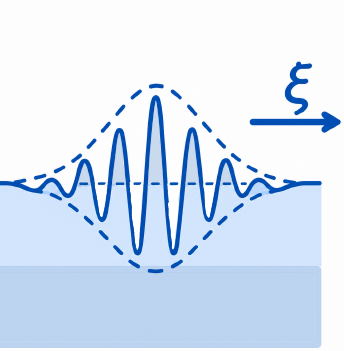}}
      & Surface-gravity waves  & $\displaystyle i\,\partial_\xi A - \partial_\tau^2 A = 0$                         & $V=-F\tau$ \\
    \br
  \end{tabular}
  \caption{\textbf{The linear potential across Schrödinger-type platforms.}
  The quantum matter-wave $\Psi$, the optical envelope $U$, and the surface-gravity
  water-wave amplitude $A$ each obey a free Schrödinger-type equation of
  identical structure, differing only in the evolution variable and the field.
  Applying a linear potential $V=-Fq$, with $F$ a constant generalized force
  along the coordinate $q$, imprints the general cubic phase dynamics
  studied here, displayed in full by its Airy eigenstates. 
  Among the three platforms, the matter-wave case alone is genuinely quantum, 
  the others being classical analogues representative of the broader class of Schrödinger-type systems. 
  Platform conventions adapted from Ref.~\cite{rozenman2019quantum}.}
  \label{tab:platforms}
\end{table*}

\section{Theoretical Model}
\label{sec:model}

The general approach considers a system accurately described, or well
approximated, by the Schrödinger equation,
\begin{equation}
  i\hbar\,\partial_t \Psi(\mathbf{r},t)
  = \Bigl[-\tfrac{\hbar^{2}}{2m}\nabla^{2} + V(\mathbf{r},t)\Bigr]\Psi(\mathbf{r},t),
  \label{eq:schroedinger}
\end{equation}
where $\Psi(\mathbf{r},t)$ is the wave function, normalized as appropriate for
the scenario considered. For convenience, a quasi-one-dimensional
description along the $x$ direction is adopted, which extends directly to
higher dimensions.
For a linear potential $V(x)=Fx$, $F$ is the gradient and $F_\mathrm{phys}=-F$ the physical force. The choice $F>0$ is a convention that reduces the rescaled equation to the canonical Airy form, with $\mathrm{Ai}$ decaying on the classically forbidden side. The quantitative figures follow it, while the schematic concept figure and Table~\ref{tab:platforms} write $V=-Fx$, there with $F$ the physical force rather than the gradient.
A stationary ansatz
$\Psi(x,t)=\psi(x)\,e^{-iEt/\hbar}$ reduces Eq.~\eqref{eq:schroedinger} to the eigenvalue
problem $[-\tfrac{\hbar^{2}}{2m}\partial_x^{2}+Fx]\,\psi=E\,\psi$.
With the eigenvalue absorbed into a shift of the origin by $\ell=E/F$, the classical
turning point where $V=E$, lengths measured from it in units of a scale $x_0$ through
$\xi=(x-\ell)/x_0$, and the gradient written as $F=f\,F_\mathrm{ref}$, where
$F_\mathrm{ref}$ is an as yet unspecified reference force, the problem takes the
homogeneous form
\begin{equation}
  -\widetilde\psi''(\xi)
  + f\,\frac{2m F_\mathrm{ref}\,x_0^{3}}{\hbar^{2}}\;\xi\,\widetilde\psi(\xi) = 0 ,
  \label{eq:rescaled_stationary}
\end{equation}
with $\widetilde\psi(\xi)=\psi(x_0\xi+\ell)$. The length $x_0$ is not fixed by the
rescaling but by the prepared profile itself, and requiring that this reduce at $f=1$ to
the canonical Airy equation $-\widetilde\psi''+\xi\,\widetilde\psi=0$ binds the reference
force to that length,
\begin{equation}
  F_\mathrm{ref}=\frac{\hbar^{2}}{2m\,x_0^{3}}\equiv F_\mathrm{eigen},
  \quad
  f=\frac{F}{F_\mathrm{eigen}},
  \quad
  E\big|_{f=1}=F_\mathrm{eigen}\,\ell .
  \label{eq:eigenforce}
\end{equation}
The \emph{eigenforce} $F_\mathrm{eigen}$ is therefore the outcome of that requirement, and
it is a property of the prepared state rather than a term in the Hamiltonian that governs
the subsequent evolution.
The same relation follows
directly from stationarity, without any rescaling: inserting
$\mathrm{Ai}\bigl(\sigma(x-\ell)/x_0\bigr)$ into the stationary equation, with
$\sigma=\pm1$ the orientation of the profile, leaves its left-hand side constant in $x$
only for $F_\mathrm{eigen}=\sigma\,\hbar^{2}/(2mx_0^{3})$. The canonical Airy form
requires the canonical orientation, so Eq.~\eqref{eq:eigenforce} is written for
$\sigma=+1$, while in general $x_0$ remains a positive length and the sign of the
eigenforce is carried by the orientation alone.

The energy is continuous and is not a distinguished discrete level, but a
label fixed by the shift $\ell$, and hence by the choice of coordinate origin rather than
solely by the state itself. The same energy is compatible with different pairs
$(F_\mathrm{eigen},\ell)$, so it is not the eigenvalue that characterizes the state, as an
eigenvalue problem would suggest, but the eigenforce, which does so uniquely up to the
shift $\ell$ and hence up to a rigid translation of the profile.
The physical content of
the state is thus the length $x_0$, equivalently $F_\mathrm{eigen}$, which with
\begin{equation}
  x_0=\Bigl(\frac{\hbar^{2}}{2mF_\mathrm{eigen}}\Bigr)^{1/3},\qquad
  t_0=\frac{2m x_0^{2}}{\hbar}
  \label{eq:scales}
\end{equation}
also fixes the time scale $t_0$. The non-dimensionalization used below is thereby
fixed~(Table~\ref{tab:scales}):
lengths and times are measured in $x_0$ and $t_0$, the dimensionless
shift is $\epsilon=\ell/x_0$, and the physical force on the packet is
$F_\mathrm{phys}=-f\,F_\mathrm{eigen}$.

Of the two solutions of the canonical Airy equation only $\mathrm{Ai}(\xi)$ remains
bounded, $\mathrm{Bi}(\xi)$ diverging on the classically forbidden side
$\xi\to+\infty$~\cite{abramowitz1964} where $\mathrm{Ai}$ decays.
Not square-integrable, the ideal Airy eigenstate is not itself a physical state, and the
admissible realization is its truncated, finite-energy form, taken here as the initial
state,
\begin{equation}
  \psi_0(\xi') = \mathcal{N}\,\mathrm{Ai}(\xi')\, e^{a\,\xi'} ,
  \qquad a > 0 ,
  \label{eq:initial_airy}
\end{equation}
with the ideal eigenstate recovered as $a\to0$~\cite{siviloglou2007ol,fu2015airy}.\footnote{%
The normalization is
$\mathcal{N}=(8\pi a)^{1/4}e^{-a^{3}/3}$ for unit probability in $\xi'$, and
$\mathcal{N}/\sqrt{x_0}$ in $x$. It vanishes as $a\to0$, the ideal eigenstate not being
square-integrable.}

Propagation in the
linear potential $V=f\xi$ is governed exactly by the Feynman
propagator~\cite{feynman1965qm_path-integrals,kennard1927,zimmermann2017},
\begin{equation}
  G(\xi,\tau;\xi',0) = \frac{1}{\sqrt{4\pi i\,\tau}}\; e^{\,i\,S_{\mathrm{cl}}} ,
  \label{eq:propagator}
\end{equation}
\begin{equation}
  S_{\mathrm{cl}}
  = \frac{(\xi-\xi')^{2}}{4\tau}
  - \frac{f(\xi+\xi')\,\tau}{2}
  - \frac{f^{2}\tau^{3}}{12} .
  \label{eq:action}
\end{equation}
Carrying out the Gaussian integral over $\xi'$ and the integral over the wavenumber $k$
of the Airy representation $\mathrm{Ai}(s)=\tfrac{1}{2\pi}\!\int\! dk \,e^{i(k^{3}/3+ks)}$ yields the evolved packet in closed form,
\footnote{For the subsequent phase analysis, the Airy argument is taken to be
real, its imaginary part $\propto a\tau$ being neglected.
This approximation is justified for sufficiently small truncation $a$ and becomes exact as $a\to0$.
The residual finite-$a$ influence along the main lobe is quantified in
Appendix~\ref{app:numerics} and discussed in the supplementary material of
Ref.~\cite{rozenman2019amplitude}.}
\begin{equation}
  \psi(\xi,\tau) = C\, e^{-\,\Im\,\theta}\, e^{\,i\,\Re\,\theta}\,
                   \mathrm{Ai}(\zeta) ,
  \label{eq:evolved}
\end{equation}
\begin{align}
  \zeta &= \xi + (f-1)\tau^{2} + 2\,i\,a\,\tau ,
        \label{eq:airy_arg}\\[2pt]
  \Re\,\theta &=
        \xi\tau - f\xi\tau + a^{2}\tau - f\epsilon\tau
        \;\underbrace{-\,\tfrac{2\tau^{3}}{3}}_{\text{intrinsic}}
        \;\underbrace{+\,f\tau^{3}\vphantom{\tfrac{2\tau^{3}}{3}}}_{\text{cross}}
        \;\underbrace{-\,\tfrac{f^{2}\tau^{3}}{3}}_{\text{induced}} ,
        \label{eq:phase}\\[2pt]
  \Im\,\theta &=
        -\,a\,\xi + 2\,a\tau^{2} - a f\tau^{2} .
        \label{eq:envelope}
\end{align}
with $C$ a normalization factor.
In the eigenstate limit $a\to0$, Eq.~\eqref{eq:evolved} describes the non-dispersing evolution of the Airy eigenstate under the linear potential, henceforth its \emph{eigendynamics}.

The cubic-in-time phase in Eq.~\eqref{eq:phase} separates into the Airy-intrinsic term $-2\tau^{3}/3$~\cite{berrybalasz1979},
the force-induced Kennard term $-f^{2}\tau^{3}/3$~\cite{kennard1927,kennard1929}, and a cross term $+f\tau^{3}$~\cite{berrybalasz1979}, falling into distinct universality classes treated in Sec.~\ref{sec:universality}.

Reading off the $\tau^{3}$ term in Eq.~\eqref{eq:phase} gives the single-packet
cubic coefficient
\begin{equation}
  c_{3,i}^{(\tau)} = \frac{-f_i^{2}+3f_i-2}{3} = -\frac{(1-f_i)(2-f_i)}{3} ,
  \label{eq:c3single}
\end{equation}
whose factorized form makes the two zeros explicit. It vanishes at $f_i=1$, where
the applied force cancels the intrinsic self-acceleration and the packet rests as
the Airy eigenstate, and at $f_i=2$, the non-trivial zero at which the cross term
offsets the intrinsic and force-induced contributions so that the cubic phase
cancels while the packet keeps accelerating.
This single-packet coefficient is the central object of the following,
its measurement through interference deferred to Sec.~\ref{sec:landscape}.

\subsection{Dimensional form}
\label{sec:realization}

Applied per packet, the scales $x_{0,i},t_{0,i}$ of Eq.~\eqref{eq:scales} follow by $F_\mathrm{eigen}\to F_{\mathrm{eigen},i}$. Restoring units in
Eq.~\eqref{eq:c3single} via $\tau_i=t/t_{0,i}$ and $f_i\to -F_{\mathrm{phys},i}/F_{\mathrm{eigen},i}$,
the single-packet cubic coefficient reads
\begin{equation}
  c_{3,i} = -\frac{(F_{\mathrm{phys},i}+F_{\mathrm{eigen},i})(F_{\mathrm{phys},i}+2F_{\mathrm{eigen},i})}{6\hbar m},
  \label{eq:c3dim}
\end{equation}
with units of $[\text{time}]^{-3}$. Its two factors place the zeros in physical
force, the eigenstate line $F_{\mathrm{phys},i}=-F_{\mathrm{eigen},i}$ and the
non-trivial line $F_{\mathrm{phys},i}=-2F_{\mathrm{eigen},i}$.

At the level of the main lobe, the global maximum of $|\psi|^{2}$, the applied force and the eigenforce act on an equal footing~\cite{berrybalasz1979,rozenman2019amplitude}: the Airy caustic, the fold at the classical turning point where neighbouring trajectories converge, self-accelerates as if driven by the eigenforce,
and the lobe, offset from the caustic by $1.019\,x_{0,i}$, the first zero of $\mathrm{Ai}'$~\cite{abramowitz1964}, in the ideal profile, accelerates at $a_\text{lobe}=(F_{\mathrm{phys},i}+F_{\mathrm{eigen},i})/m$.
The plus sign is that of a Newtonian force balance, so the two oppose one another where their signs differ and reinforce one another where they agree. The two factors weight the eigenforce differently. In the first, $F_{\mathrm{phys},i}+F_{\mathrm{eigen},i}$, it is the effective total force on the lobe and enters with unit weight. In the second, $F_{\mathrm{phys},i}+2F_{\mathrm{eigen},i}$, it enters with twice the weight. Each factor vanishes where the applied force cancels the eigenforce at its respective weight. The caustic--lobe offset and the lobe acceleration are exact for the ideal eigenstate. For the finite-energy packet of Eq.~\eqref{eq:initial_airy} the truncation shifts the offset and renders it time dependent, and the profile stays quasi-invariant only until diffraction takes over~\cite{siviloglou2007prl}, so that both hold over a finite interval and become exact as $a\to0$.

The eigenforce thus enters the cubic phase on the same algebraic footing as the applied force, and the coefficient vanishes precisely where the two balance. In this sense it acts as an antagonist to the applied force, the statement referring to that force balance. This involves no conflict with any conservation law: the antagonism is effective, not real, the eigenforce belonging to the state and not to the Hamiltonian.
Ehrenfest's theorem constrains the centroid $\langle x\rangle=\int x|\psi|^{2}\,\mathrm{d}x$ of the normalizable packet alone,
\begin{equation}
  a_\text{lobe}=\frac{F_{\mathrm{phys},i}+F_{\mathrm{eigen},i}}{m},
  \qquad
  a_{\langle x\rangle}=\frac{F_{\mathrm{phys},i}}{m},
  \label{eq:lobe_vs_centroid}
\end{equation}
the eigenforce entering the one and absent from the other. An effective force can thus oppose the applied one in the caustic's self-acceleration and within the phase alike, with no counterpart in the centroid motion.

\begin{table}[t]
  \centering
  \begin{tabular}{lll}
    \toprule
    Quantity & Dimensionless & Physical scale \\
    \mr
    Length & $\xi=(x-\ell)/x_0$ & $x_0=(\hbar^{2}/2mF_\mathrm{eigen})^{1/3}$ \\
    Time   & $\tau=t/t_0$    & $t_0=2m x_0^{2}/\hbar$ \\
    Force  & $f$                & $F_\mathrm{phys}=-f\,F_\mathrm{eigen}$ \\
    \br
  \end{tabular}
  \caption{Mapping of the dimensionless variables to physical units. For a
  dilute BEC, $m$ is the atomic mass and $F_\mathrm{eigen}$ the eigenforce,
  the same rescaling carries over to other Schrödinger-analogue platforms (e.g.\
  surface-gravity water waves~\cite{rozenman2019amplitude}) with $m$, $F_\mathrm{eigen}$
  replaced by the corresponding effective parameters.}
  \label{tab:scales}
\end{table}

\subsection{Cubic-phase landscape and signal optimization}
\label{sec:landscape}

\begin{figure}[t]
  \centering
  \includegraphics[width=\columnwidth,trim=295 98 220 160,clip]{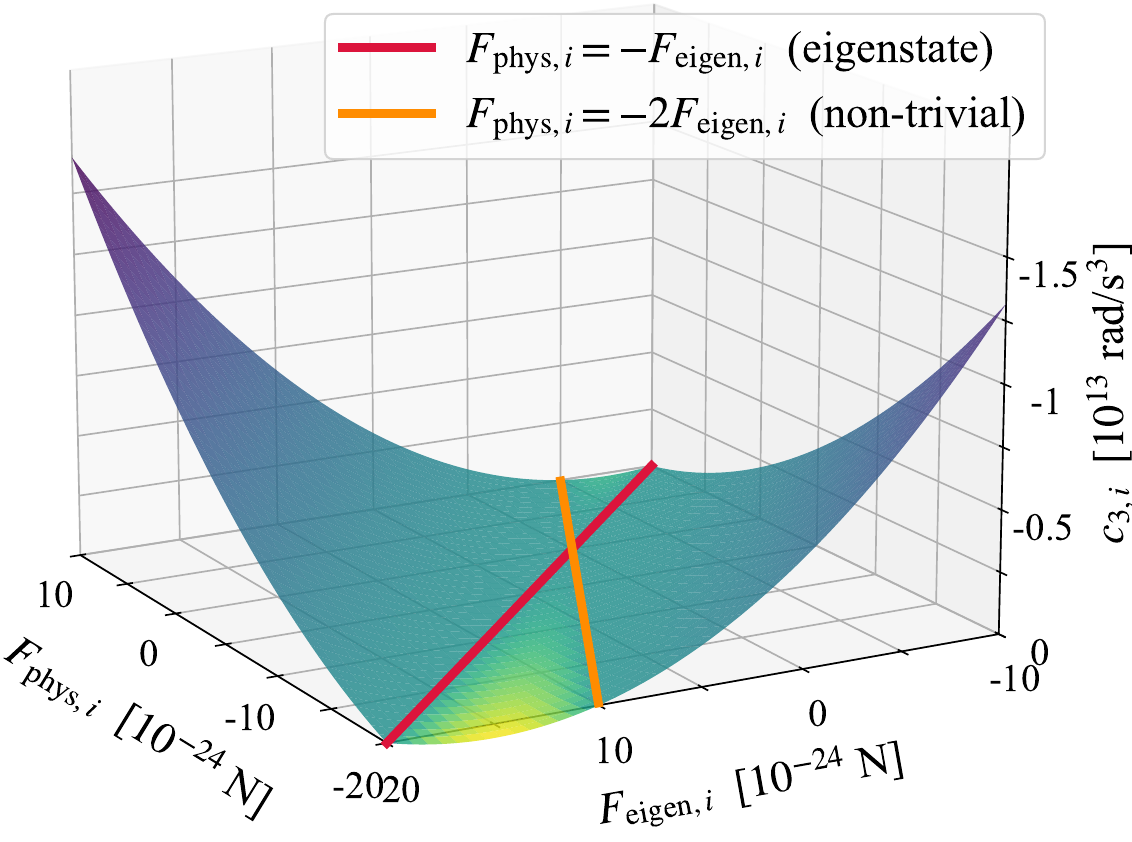}
  \caption{\textbf{Cubic-phase landscape.}
  Single-packet cubic coefficient $c_{3,i}$, Eq.~\eqref{eq:c3dim}, as a function
  of the applied force $F_{\mathrm{phys},i}$ and the eigenforce $F_{\mathrm{eigen},i}$, for
  $^{87}$Rb with $m_{^{87}\mathrm{Rb}}\,g$ setting the force scale.
  The indefinite quadratic form yields a saddle surface that vanishes
  along the eigenstate line $F_{\mathrm{phys},i}=-F_{\mathrm{eigen},i}$ (red) and the
  non-trivial line $F_{\mathrm{phys},i}=-2F_{\mathrm{eigen},i}$ (orange), and grows
  quadratically in magnitude along any ray from the origin other than those two lines,
  marking the regime of strong cubic signature. Both signs of $F_{\mathrm{eigen},i}$ are shown, the
  surface being symmetric under
  $(F_{\mathrm{phys},i},F_{\mathrm{eigen},i})\to(-F_{\mathrm{phys},i},-F_{\mathrm{eigen},i})$
  since the two orientations of the Airy profile are related by parity. Along
  $F_{\mathrm{eigen},i}=0$ the eigenstate-based scaling is singular, $x_{0,i}$ diverging,
  while $c_{3,i}$ itself stays regular and reduces, at fixed $F_{\mathrm{phys},i}$, to the
  universal force-induced term. Colour
  encodes the same $c_{3,i}$ as the vertical axis and carries no independent
  information. Up to a
  constant offset set by the reference packet, the same landscape gives the relative
  cubic coefficient~(Eq.~\ref{eq:c3rel}) accessed in interference.}
  \label{fig:landscape}
\end{figure}

Viewed as a function of the eigenforce and the applied force, the single-packet
cubic coefficient~(Eq.~\ref{eq:c3dim}) forms a quadratic landscape, shown in
Fig.~\ref{fig:landscape}.
Its two zero lines cross at the origin and divide the plane into
regions of opposite sign. Being a homogeneous quadratic form, $c_3$ scales
quadratically under a radial rescaling at fixed direction, at a rate that depends on
the direction and vanishes along the zero lines, where the coefficient is zero at any
distance from the origin. A pronounced cubic signature is therefore
found away from both zero lines, which identifies the force regime most favourable
for resolving the phase. For a given bound on the realizable forces it is largest in
the quadrants where $F_\mathrm{phys}$
and $F_\mathrm{eigen}$ share the same sign, for which the cross term $3F_\mathrm{eigen}F_\mathrm{phys}$ adds
constructively to the always-positive $F_\mathrm{phys}^2$ and $2F_\mathrm{eigen}^2$ contributions.

Both signs of $F_\mathrm{eigen}$ occur in Fig.~\ref{fig:landscape}. The derivation is
carried out for the canonical orientation $\sigma=+1$ and carries over to $\sigma=-1$ by
parity: under $x\to-x$ the kinetic term is unchanged while $Fx$ reverses sign, so a packet
of reversed orientation in a given potential is the original packet in the sign-reversed
one. Being spatially uniform, the cubic phase is carried over unchanged, and
Eq.~\eqref{eq:c3dim} holds with $F_\mathrm{eigen}\to-F_\mathrm{eigen}$. The landscape is
accordingly symmetric under
$(F_\mathrm{phys},F_\mathrm{eigen})\to(-F_\mathrm{phys},-F_\mathrm{eigen})$, both zero
lines mapping onto themselves. The parametrization itself is defined for $x_0>0$ together
with the orientation $\sigma$, and the eigenstate-based scaling is singular as
$F_\mathrm{eigen}\to0$, where $x_0$ and $t_0$ diverge. The coefficient, however, stays
regular: at fixed $F_\mathrm{phys}$ it reduces to $-F_\mathrm{phys}^{2}/(6\hbar m)$, which
Sec.~\ref{sec:universality} identifies as the universal, shape-independent contribution.
The $F_\mathrm{eigen}=0$ cross-section of the landscape is thus the part of the cubic
phase shared by every wave packet in a linear potential.

A spatially uniform cubic phase is a global phase, unobservable for a single
packet, and becomes accessible only as the relative phase of two colliding
packets,
\begin{equation}
  \Delta\theta = \Re\,\theta^{(1)} - \Re\,\theta^{(2)} .
  \label{eq:relphase}
\end{equation}
Each packet $i$ is then governed by its own linear potential, hence its own force
$F_{\mathrm{phys},i}$, realized for instance through spin-dependent magnetic
gradients~\cite{amit2019,zimmermann2017,koehl2001}. In the laboratory frame, each packet
launched from $\ell_i$,\footnote{The constant-in-position term
$-\ell_i/(t_{0,i}x_{0,i})$ in $c_1$ originates from the global phase
$-f_i\epsilon_i\tau_i$ of the shifted linear potential, with
$\epsilon_i=\ell_i/x_{0,i}$. Being spatially uniform, it leaves the single-packet
density unchanged, yet for two distinctly prepared packets it does not cancel and
enters the measurable relative phase, so it is not a discardable constant in the
general Airy collision.} the relative phase takes the form
$\Delta\theta=c_3 t^{3}+c_1 t$ with $c_2=c_0=0$, purely cubic and linear in time,
a cleanly separable cubic signature. Its linear coefficient
\begin{equation}
  c_1 = \frac{(1-f_1)\,x - \ell_1}{t_{0,1} x_{0,1}}
      - \frac{(1-f_2)\,x - \ell_2}{t_{0,2} x_{0,2}}
      + \frac{a_1^{2}}{t_{0,1}} - \frac{a_2^{2}}{t_{0,2}}
  \label{eq:c1}
\end{equation}
sets the spatial fringe spacing, while the cubic coefficient is the difference of
the two single-packet contributions~(Eq.~\ref{eq:c3dim}), the \emph{relative cubic
coefficient},
\begin{equation}
  c_3 = \frac{g(F_{\mathrm{eigen},2},F_{\mathrm{phys},2})
              -g(F_{\mathrm{eigen},1},F_{\mathrm{phys},1})}{6m\hbar},
  \label{eq:c3rel}
\end{equation}
with $g(F_\mathrm{eigen},F_\mathrm{phys})=(F_\mathrm{eigen}+F_\mathrm{phys})(2F_\mathrm{eigen}+F_\mathrm{phys})$.
The relative coefficient is odd under exchange of the two packets and even under parity, so it vanishes for any preparation left invariant by the exchange, alone or combined with parity.
Since each packet enters through the same combination $g$, fixing one packet
shifts the contribution of the other by a constant. The relative coefficient
thus inherits the single-packet structure, and a separate landscape is not
required. The constant offset is system-dependent, set by the reference packet, but
it translates only the zero level and leaves the landscape shape, and with it
the physics, unchanged. What changes is the location of the zero point, as the relative phase
vanishes whenever the two contributions coincide, $g_1=g_2$, rather than on the
individual lines. A strong relative signal accordingly demands that the two single-packet
contributions $g_1,g_2$ differ as much as possible. This is
realized when they occupy regions of opposite sign of the cubic landscape, so
that $g_1$ and $g_2$ enter with opposite signs and their difference accumulates
rather than partially cancels.
Of this general relative form, the experimentally realized closed $T^3$-interferometer phase~\cite{zimmermann2017,amit2019} is a particular initial-state-independent case, and single-packet waveform measurements in an analogue system~\cite{rozenman2019amplitude} a per-packet counterpart.

In the experimentally common case of a shared potential, both packets are
subject to the same applied force $F_\mathrm{phys}$, and the relative
coefficient factorizes,
\begin{equation}
\begin{aligned}
  c_3=\frac{1}{3m\hbar}\,&(F_{\mathrm{eigen},2}-F_{\mathrm{eigen},1})\\
      &\, \cdot \,(F_{\mathrm{eigen},1}+F_{\mathrm{eigen},2}+\tfrac{3}{2}F_\mathrm{phys}).
\end{aligned}
  \label{eq:c3common}
\end{equation}
The $F_\mathrm{phys}^{2}$ contributions cancel between the packets, so that the
cubic signal vanishes either when the two eigenforces coincide,
$F_{\mathrm{eigen},1}=F_{\mathrm{eigen},2}$, or at the non-trivial zero where
the total-force factor vanishes, $F_\mathrm{phys}=-\tfrac{2}{3}(F_{\mathrm{eigen},1}+F_{\mathrm{eigen},2})$,
and otherwise scales with the product of the two factors. Both branches together
constitute the factorized equal-force form of the general zero condition $g_1=g_2$. The design
criterion thus becomes concrete, as a strong relative coefficient demands
pairing two distinctly prepared packets in a sufficiently strong common
potential.

The objective $|c_3|$ is itself unbounded and is capped only by the realizable
forces. The experimentally decisive quantity, however, is the precision of the
extracted coefficient, which trades the growth of $c_3$ against the shorter
fringes and hence reduced fit windows at stronger forces, an interior optimum that,
for generalized states and realistic noise, no longer admits a closed form.

The same contrast principle extends to non-Airy packets: the relative coefficient
remains the difference of single-packet contributions, with the intrinsic scale
now a chosen length, typically the packet width, rather than the eigenforce-fixed Airy length $x_0$, so that, e.g., two Gaussians
acquire a strong relative cubic coefficient through distinct applied forces or
sufficiently differing widths.

\subsection{Universality across wave-packet shapes and platforms}
\label{sec:universality}

Because the classical action~(Eq.~\ref{eq:action})~\cite{pellner2026thesis} is quadratic in the force, no higher powers of $F_\mathrm{phys}$ arise, and the single-packet cubic coefficient~(Eq.~\ref{eq:c3dim}) comprises exactly three contributions falling into distinct universality classes,
\begin{equation}
  c_{3,i} = \underbrace{-\frac{F_{\mathrm{phys},i}^{2}}{6\hbar m}}_{\text{force-induced}}
  \;\underbrace{-\,\frac{F_{\mathrm{eigen},i}F_{\mathrm{phys},i}}{2\hbar m}}_{\text{cross}}
  \;\underbrace{-\,\frac{F_{\mathrm{eigen},i}^{2}}{3\hbar m}}_{\text{intrinsic}} .
  \label{eq:universality_decomposition}
\end{equation}
The force-induced term is universal: it is a spatially uniform, global phase whose cubic coefficient $-F_\mathrm{phys}^{2}/(6\hbar m)$~\cite{kennard1927,kennard1929} is common to every wave packet in a linear potential~\cite{zimmermann2017}. Being independent of the initial coordinate, the force term in the classical action factors out of the evolution integral as a global phase irrespective of the initial profile, extending to any Schrödinger-type platform. The Airy eigenstate and a stationary Gaussian, which differ in all shape-dependent terms, share it exactly. The shape enters through the intrinsic and the cross terms, the intrinsic being the $-F_\mathrm{eigen}^{2}/(3\hbar m)$ of the dispersionless Airy eigenstate while reducing to the width-dependent Gouy phase for a Gaussian~\cite{fengwinful2001,rozenman2021projectile}, and the cross term $-F_\mathrm{eigen}F_\mathrm{phys}/(2\hbar m)$ the shape-dependent coupling of the applied force to the eigenforce behind the packet's self-acceleration, vanishing in its absence. For a general, dispersing packet these shape-dependent contributions lose their constant-coefficient $t^{3}$ form, their coefficients turning time-dependent as the packet spreads. The Airy functions, being the eigenstates of the linear potential, are thereby form-preserving carriers of this structure,
and their interference exposes the phase in its cleanest, non-dispersing setting. Across platforms the analysis carries over unchanged, as every Schrödinger-type field obeys the same evolution. Across wave-packet shapes, only the universal force-induced term is shared, so the Airy eigenstates are the canonical realization in which these dynamics take a closed, effective Newtonian form.

\section{Extraction algorithms}
\label{sec:extraction}

A quantum measurement returns the density alone, so a spatially uniform phase
such as the cubic-in-time term is accessible only as a \emph{relative} phase
between interfering components. Colliding two packets renders it observable: the
superposition density
\begin{equation}
  \rho = |\psi_1|^{2} + |\psi_2|^{2} + 2\,|\psi_1|\,|\psi_2|\,\cos\Delta\varphi
  \label{eq:density}
\end{equation}
carries $\Delta\varphi=\arg\psi_1-\arg\psi_2$ in the fringe term, whose
cubic-in-time part is the target quantity. Experimentally, the observable is the
density itself, recorded as a time series of interference images: absorption
images of the atomic cloud for a Bose--Einstein condensate, and surface-elevation
or intensity profiles for the water-wave and optical realizations. At each time
the spatial fringe fixes $\Delta\varphi(\tau)$, whose evolution in the running
variable $\tau$ yields the cubic coefficient $c_3$.

The procedure, its systematics and its technical details are given in full in
Ref.~\cite{pellner2026phase}. Here it resolves the full coefficient of two packets under
distinct applied forces, force-induced and cross term included.
The relative phase is isolated by demodulation~\cite{takeda1982} over a region of interest centred
at $\xi_\mathrm{c}(\tau)$ on the interference fringe of the main lobes, usually the most intense among the
pattern's temporally contiguous fringes. A Gaussian filter $w(\xi-\xi_\mathrm{c}(\tau))$,
together with removal of the differential carrier $k_\mathrm{f}(\tau)$ fixed beforehand from
single-packet calibration, gives the complex signal
\begin{equation}
  S(\tau) = \int w(\xi-\xi_\mathrm{c}(\tau))\,\psi_2^{*}(\xi,\tau)\,\psi_1(\xi,\tau)\,
            e^{-\mathrm{i}k_\mathrm{f}(\tau)[\xi-\xi_\mathrm{c}(\tau)]}\,\mathrm{d}\xi .
  \label{eq:hpe_signal}
\end{equation}
Only the carrier is removed by the
integration, while the cubic-in-time phase, uniform in $\xi$, passes through it unchanged
and is read off $\arg S$, from which a low-order polynomial fit over a
connected fringe segment returns $c_3$.
The demodulation, however, references a
time-dependent and only approximately known carrier $k_\mathrm{f}(\tau)$ and fringe centre
$\xi_\mathrm{c}(\tau)$, together with a local plane-wave approximation, and the resulting
low-order-in-time bias leaves only the cubic coefficient as the physically meaningful
output.

Two complementary methods can be employed. The heterodyne scheme is the more precise and
underlies all results reported here: it reconstructs the coherence term
$\psi_2^{*}\psi_1$, in an experiment by phase-shifting interferometry~\cite{bruning1974}
from density images at controlled relative-phase offsets. The
density-based scheme operates on the measured density alone~\cite{takeda1982}, subtracting a
filtered local background, and trades precision for experimental simplicity. Both
rest on the same assumptions: an interference amplitude varying slowly on the
fringe scale, a locally plane-wave carrier, and a quasi-stationary fringe
position.
Because the fit lies where the cubic term runs against the dominant lower-order
trend, $c_3$ is sensitive to the fit window, and a symmetric selection around the
inflection point avoids an asymmetry bias~\cite{pellner2026phase}. The window is
calibrated: at a given parameter configuration it is the one whose fit reproduces the
analytically known coefficient most closely, which debiases the estimator there and
yields the window to be applied to measured data, where the coefficient is not known in
advance. The sensitivity of the estimate to this choice is characterized in
Ref.~\cite{pellner2026phase} and constitutes a dominant systematic of the
extraction. The quoted
uncertainties are the ordinary least-squares standard errors of the fit.

\section{Numerical demonstration}
\label{sec:demonstration}

\begin{figure*}[!tb]
  \centering
  \includegraphics[width=\textwidth,trim=5 0 0 0,clip]{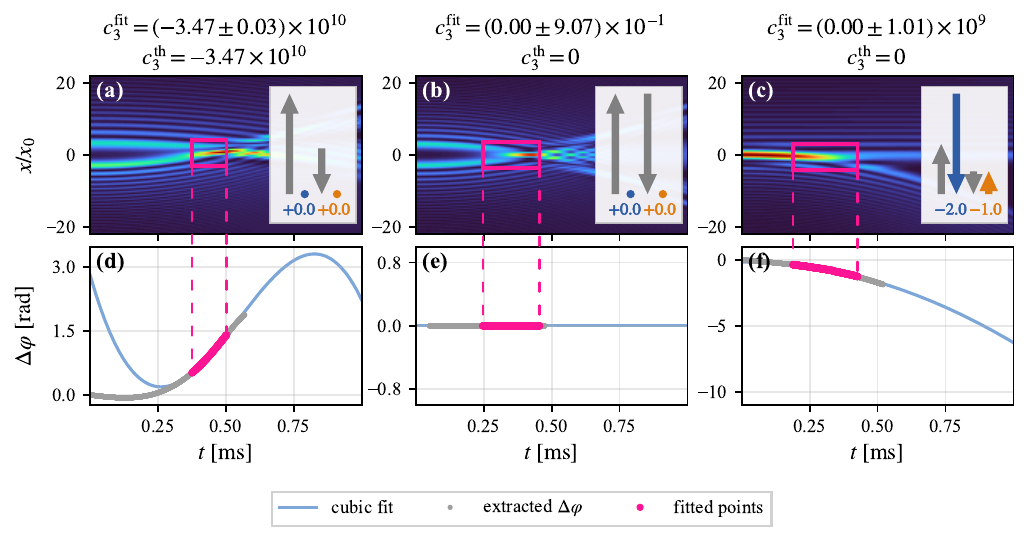}
  \caption{\textbf{Relative cubic phase across collision regimes.} Three further
  simulated Airy--Airy collisions ($^{87}$Rb) in the linear potential $V=+Fx$, each column a collision regime: the
  density evolution (top row, \emph{(a)}--\emph{(c)}, with the main interference
  fringe boxed and linked to the fit window) and the relative phase
  $\Delta\varphi(t)$ with its cubic least-squares fit (bottom row,
  \emph{(d)}--\emph{(f)}). The density panels are shown qualitatively in colour, as in Fig.~\ref{fig:hero}.
  Fitted and theoretical coefficients $c_3^{\mathrm{fit}}$ and
  $c_3^{\mathrm{th}}$ are quoted per panel in $\mathrm{rad\,s^{-3}}$, with the
  power of ten shown alongside each. As in
  Fig.~\ref{fig:hero}, phase points beyond the first fringe are omitted~\cite{note_phase_cutoff}. Each density panel
  carries the per-packet force inset of Fig.~\ref{fig:hero}, the coloured arrow the
  applied force and the grey arrow the eigenforce, on a common scale and signed by
  the arrowhead end. \emph{(a,d)} Scale asymmetry: two
  freely propagating packets of unequal width
  ($F_\mathrm{phys}=0$, $\beta_2/\beta_1=1.3$), where the signal originates solely from
  the difference in scale. \emph{(b,e)} Symmetric cancellation: mirror-symmetric free packets
  ($F_\mathrm{phys}=0$, $\beta_1=\beta_2$) carry equal cubic coefficients, so their
  difference vanishes. \emph{(c,f)}
  Non-trivial cancellation: one packet on the eigenstate line
  $F_\mathrm{phys}=-F_\mathrm{eigen}$ (static) and one on the non-trivial line
  $F_\mathrm{phys}=-2F_\mathrm{eigen}$ (accelerating), at unequal width ($\beta_2/\beta_1=1.3$),
  starting with coincident main lobes rather than separated,
  both with vanishing single-packet coefficients for distinct reasons, again yielding $c_3=0$.
  In both cancellation regimes the fitted cubic coefficient vanishes within its
  uncertainty, and the late-time curvature of the plotted best-fit merely
  reflects this finite point estimate, extrapolated beyond the fit window,
  rather than a resolved cubic signal.}
  \label{fig:scenarios}
\end{figure*}

The framework is verified on a representative Airy--Airy collision~(Fig.~\ref{fig:hero}) propagated by
a split-step Fourier integrator~\cite{javanainen2006}~(Appendix~\ref{app:numerics}). The atomic mass and the
gravitational force scale $mg$, the only dimensional inputs, are those of
$^{87}$Rb and set the collision timescale and the magnitude of $c_3$. The
underlying dynamics are system-independent. One packet is prepared as an Airy eigenstate whose eigenforce is set to $mg$, rendering it approximately stationary
over the simulated interval and exact as $a\to0$. Its cubic coefficient vanishes identically in the same limit, so the relative
cubic coefficient reduces to that of the second, accelerating packet, which isolates the
predicted coefficient of the accelerating packet and makes the collision a clean setting
for its extraction.

The relative phase is recovered by heterodyne demodulation over the main interference fringe.
A cubic fit over the connected fringe segment, with the window fixed as described in
Sec.~\ref{sec:extraction}, returns
$c_3=(-1.662\pm0.021)\times10^{11}\,\mathrm{rad\,s^{-3}}$ against the predicted
$-1.662\times10^{11}\,\mathrm{rad\,s^{-3}}$, the calibrated estimator
reproducing it with a deviation of $0.01\%$ at this parameter configuration, the quoted
uncertainty being the standard error of the fit.\footnote{The calibration debiases the estimator at this
parameter choice, so the $0.01\%$ is the deviation of the debiased estimator there rather
than a bound on its bias at other configurations. No noise model is imposed, since
realistic noise is
platform- and system-dependent. The fit residuals are accordingly not statistical scatter
but systematic model misspecification, so the standard error measures lack of fit rather
than sampling variability.
Both the quoted deviation and the standard error are set by the
numerical and algorithmic implementation at the resolution used here, about $3200$ grid points per $x_0$ and a time step
$\Delta t\approx23\,\mathrm{ns}$, with the other scenarios of the same order~(Appendix~\ref{app:numerics}), and
constitute a lower bound on any experimental uncertainty, which would require a
platform-specific noise and imaging model.}

Beyond this single configuration, the relative cubic coefficient organizes the
collision into distinct regimes (Fig.~\ref{fig:scenarios}). Across all regimes
the fitted coefficient reproduces the predicted value within one standard error.

\section{Conclusion}

A wave packet in a linear potential accumulates a cubic-in-time phase that, since the
classical action is quadratic in the applied force, splits into intrinsic, force-induced,
and cross contributions. Derived from the eigenvalue problem, the eigenforce characterizes
the Airy eigenstate uniquely up to a rigid translation, and the cubic coefficient
factorizes into two effective Newtonian force balances, the eigenforce entering at unit and
double weight. Over the plane of the two forces the coefficient is physically transparent,
vanishing under two distinct conditions -- at the eigenstate, where the packet comes to
rest, and at a non-trivial zero, where the cubic phase cancels even as the packet keeps
accelerating, demonstrated here in a simulated collision (Fig.~\ref{fig:scenarios}). The
eigenforce, itself an effective force, thereby acts as an antagonist to the applied force:
beyond the familiar opposition in the caustic's self-acceleration~\cite{berrybalasz1979},
it counters the applied force within the phase, while by Ehrenfest's theorem the centroid
responds to the physical force alone.

Throughout, the Airy eigenstates are the canonical realization. Shape enters only through
the eigenforce, so the coefficient at $F_\mathrm{eigen}=0$ is universal
across wave shapes.
In the interference of two packets, each under its own potential, this
cubic phase, individually invisible because it is spatially uniform, becomes a directly
measurable and designable observable, its magnitude the difference between the two packets'
cubic dynamics, which vanishes by symmetry when they coincide.

Resting solely on the structure of the Schrödinger equation, the cubic phase dynamics
characterized here are universal across Schrödinger-type platforms. Since any smooth
potential is linear to first order, these dynamics are the leading, exactly-solvable
contribution for arbitrary potentials, while its curvature adds further corrections. The
analysis extends directly to higher dimensions by factoring Gaussian envelopes along the
transverse directions, albeit with added experimental demands. Adding interactions carries
it beyond the linear regime, where the cubic coefficient could itself serve as a probe of
the nonlinearity, as for interacting condensates~\cite{pellner2026phase}.

Special cases have been measured in analogue and
matter-wave systems~\cite{rozenman2019amplitude,amit2019,rozenman2021projectile}. The
general cubic phase remains to be probed directly. Promising near-term candidates are
optical platforms, whose scales are widely tunable, whereas matter waves give access to a
genuine constant force, down to the microgravity regime in space and on sounding-rocket
platforms~\cite{lachmann2021space,piest2025microgravity}.

This formulation provides a platform-independent probe of the cubic phase in the linear
potential and, through its force dependence, of the constant force.

\section*{Acknowledgments}

M.L.D.D.P.\ thanks J.\,W.\,M.\ Bush for hosting this work in his research group at MIT and for valuable discussions and advice, J.\ Been for a critical reading of the derivations, and T.\ Esslinger for a helpful comment on Airy interference in ultracold atoms.
M.L.D.D.P.\ was supported as a visiting student researcher at MIT through the MIT School of Science Research Innovation Seed Fund held by G.G.R.
In addition, G.G.R.\ acknowledges support from the C.L.E.\ Moore Instructorship in Applied Mathematics.
Generative-AI tools were used for copyediting, code development, the concept schematic in Fig.~\ref{fig:concept}, and the platform icons in Table~\ref{tab:platforms}.

\section*{Author Declarations}

\subsection*{Conflict of Interest}
The authors have no conflicts to disclose.

\subsection*{Author Contributions}
M.L.D.D.P.\ developed the theory, performed the simulations and analysis, created the figures, and wrote the manuscript. G.G.R.\ supervised the project, proposed the concept figure, and revised the manuscript.

\section*{Data Availability}
Data supporting the findings of this study are available from the corresponding
author upon reasonable request.

\begingroup
\footnotesize
\setlength{\bibsep}{0pt}
\setlength{\itemsep}{0pt}
\setlength{\parskip}{0pt}
\setlength{\bibhang}{1.2em}

\bibliographystyle{unsrt}
\bibliography{references}

\begin{thebibliography}{10}
\bibitem{note_phase_cutoff}
Past the first interference fringe the relative phase is position-dependent
  through its sub-cubic-order terms, which the demodulation approximates with a
  local carrier that differs between fringe locations, so robust concatenation
  of the per-fringe fragments lies beyond the current reach of the extraction
  algorithm. Points from the first discontinuity exceeding $0.3\,\mathrm{rad}$
  onward are therefore not plotted, as are points past a smaller step at a
  fringe hand-over. The cubic fit uses only the contiguous first fringe,
  leaving $c_3$ unaffected.
\end{thebibliography}
\endgroup

\appendix
\clearpage
\balance
\section{Simulation parameters}
\label{app:numerics}

Both packets are propagated on a periodic grid by the symmetric
$V/2$--$K$--$V/2$ split-step scheme of Ref.~\cite{javanainen2006}, the kinetic
half-step taken in Fourier space, on a box of $326\,x_0$ closed by an absorbing
margin. All runs use $3.2\times10^{3}$ grid points per $x_0$ and sample
$1\,\mathrm{ms}$ of propagation at $4000$ snapshots, the time step of each run
being listed in Table~\ref{tab:runparams}. The atomic mass and the force
scale are those of $^{87}$Rb with $F_\mathrm{eigen}=mg$, which fixes $x_0$ and
$t_0$ through Eq.~\eqref{eq:scales}; the truncation is $a=0.05$ and the Gaussian
filter of the region of interest has width $0.01\,x_0$. Packet~1 carries the
orientation $\sigma_1=+1$ and packet~2 the mirrored $\sigma_2=-1$, so that
$F_{\mathrm{eigen},i}=\sigma_i F_\mathrm{eigen}/\beta_i^{3}$ and
$x_{0,i}=\beta_i x_0$, with $\beta_1=1$ throughout.

\begin{table}[h]
  \centering
  \footnotesize
  \setlength{\tabcolsep}{6pt}
  \renewcommand{\arraystretch}{1.25}
  \begin{tabular}{@{}lcccccc@{}}
    \toprule
      & & \multicolumn{2}{c}{$F_{\mathrm{phys},i}/F_{\mathrm{eigen},i}$}
      & lobe & $\Delta t$ & fit window \\
    \cmidrule(lr){3-4}
      & $\beta_2$ & $i=1$ & $i=2$ & offsets $(x_0)$ & (ns) & (ms) \\
    \mr
    Fig.~\ref{fig:hero}
      & $1.3$ & $+1.3$ & $-1$ & $\mp2$ & $23.4$ & $0.295$--$0.449$ \\
    \ref{fig:scenarios}(a,d)
      & $1.3$ & $0$ & $0$ & $\mp2$ & $53.8$ & $0.375$--$0.502$ \\
    \ref{fig:scenarios}(b,e)
      & $1.0$ & $0$ & $0$ & $\mp2$ & $53.8$ & $0.246$--$0.455$ \\
    \ref{fig:scenarios}(c,f)
      & $1.3$ & $-2$ & $-1$ & coincident & $15.2$ & $0.189$--$0.426$ \\
    \br
  \end{tabular}
  \caption{Parameters that differ between the four simulated collisions. The
  remaining settings are given in the text, the fitted coefficients in the
  captions of Figs.~\ref{fig:hero} and~\ref{fig:scenarios}.}
  \label{tab:runparams}
\end{table}

The finite-truncation residual is quantified on the stationary reference
packet~(Table~\ref{tab:truncation}), whose ideal counterpart has no displacement,
no broadening, and a vanishing cubic coefficient. The estimator
(Figs.~\ref{fig:hero} and~\ref{fig:scenarios}) is calibrated against the ideal
reference~\cite{pellner2026phase}, so this residual is absorbed in the
calibration.

\begin{table}[h]
  \centering
  \footnotesize
  \setlength{\tabcolsep}{6pt}
  \renewcommand{\arraystretch}{1.25}
  \begin{tabular}{@{}lccc@{}}
    \toprule
    $a$ & drift & lobe width & $c_3$ \\
        & $(x_0)$ & growth & $(10^{3}\,\mathrm{rad\,s^{-3}})$ \\
    \mr
    $0.05$     & $0.0067$  & $0.35\%$  & $-645\pm1203$ \\
    $0.025$    & $0.0017$  & $0.089\%$ & $-72\pm230$ \\
    $0.0125$   & $0.00043$ & $0.023\%$ & $-8\pm63$ \\
    $0.00625$  & $0.00011$ & $0.006\%$ & $-2\pm14$ \\
    $0.003125$ & $0.00003$ & $0.002\%$ & $-0.1\pm9.4$ \\
    \br
  \end{tabular}
  \caption{Finite-truncation residuals of the stationary reference packet over
  $0.35\,\mathrm{ms}$ of propagation, each entry being the residual itself. The
  drift follows the applied force. The coefficient $c_3$ is the median over the
  evaluation positions and its uncertainty a bootstrap over them, consistent
  with zero throughout. At $a=0.05$ and $20$ grid points per $x_0$, halving or
  doubling the spatial or temporal step, or doubling the domain, changes $c_3$
  by less than $0.01\%$.}
  \label{tab:truncation}
\end{table}

\end{document}